\documentclass[twocolumn,%
]{revtex4-1}

\usepackage{amsfonts,amsmath,amssymb}
\usepackage{graphicx}
\usepackage{braket}

\newcommand{\tr}[0]{\text{tr}}
\newcommand{\hilb}[0]{\mathcal{H}}
\newcommand{\midi}[0]{\mathcal{M}}

\begin{document}

\title{Quantum Correlations from Classically Correlated States}

\author{
G. Bellomo$^1$, A.P. Majtey$^{2}$,  A.R. Plastino$^{3}$, A.
Plastino$^1$}

\affiliation{$^1$IFLP-CCT-CONICET, National University La Plata\\C. C. 727, 1900 La Plata, Argentina \\
$^2$Instituto de F\'{i}sica, Universidade Federal do Rio de Janeiro, 21.942-972, Rio de Janeiro (RJ) Brazil \\
$^3$CeBio y Secretaria de Investigacion, Universidad Nacional del Noroeste de la Prov. de Buenos Aires - UNNOBA and CONICET, R. Saenz Pe\~na 456, Junin, Argentina}

\date{\today}

\begin{abstract}
Consider a bipartite quantum system with at least one of its two components being itself a composite system. By tracing over part of one (or both) of these two subsystems it is possible to obtain a reduced (separable) state that exhibits quantum correlations even if the original state of the full system is endowed only with classical correlations. This effect, first pointed out by Li and Luo in [PRA \textbf{78}, 024303 (2008)], is of considerable interest because there is a growing body of evidence suggesting that quantum correlations in non-entangled, mixed states may constitute a useful resource to implement non trivial information related tasks. Here we conduct a systematic exploration of the aforementioned effect for particular families of states of quantum systems of low dimensionality (three qubits states). In order to assess the non-classicality of the correlations of the reduced states we use an indicator of quantum correlations based upon the state disturbances generated by the measurement of local observables. We show, for a three-qubit system, that there exists a relationship between the classical mutual information  of the original classically correlated states and the maximum quantum correlation exhibited by the reduced states.

\vskip 0.5cm


\end{abstract}

\maketitle

\section{Introduction}
It has been realized in recent years that the quantum features of the correlations exhibited by multipartite quantum systems are manyfold, entanglement being only one of the possible non-classical manifestations~\cite{OZ01,HV01,Vedral03,LC10,Luo08,Luo08b,LL08,MV12,GPA11,LLZ07,DVB10,RRA2011,MPP12,SRN08,RS10,WPM09,AD10,OHHH02,LF11,RCC10,SAPB12}. Even separable mixed states (that are not entangled) can have correlations with subtle non-classical properties~\cite{OZ01}.
Several quantitative measures have been proposed to study the different non-classical aspects (besides entanglement) of the correlations appearing in composite quantum systems. Among these measures we can mention quantum discord~\cite{OZ01} and the measures of correlations based on the disturbances of quantum states arising from local measurements. The latter ones were advanced by Luo~\cite{Luo08,Luo08b,LL08} and by SaiToh and collaborators~\cite{SRN08,RS10}. In the case of pure states all these measures coincide with quantum entanglement. However, in the case of mixed states these quantities correspond to physical properties of quantum systems that differ from entanglement.

It is generally agreed that the states of a bipartite system that are to be regarded as being only classically correlated (that is, having no quantum correlations) are those described by density matrices that are diagonal in a product basis ${\{\ket{i}\ket{j},\,i=1,\ldots,N_1;\, j=1,\ldots, N_2\}}$, where ${\{\ket{i},\,i=1,\ldots,N_1\}}$ and ${\{\ket{j},\,j=1,\ldots,N_2\}}$ are orthonormal bases associated with the two subsystems, ${N_{1,2}}$ being the dimensions of the concomitant Hilbert spaces. It is worth stressing that the set of classical states is different from the set of separable (that is, non entangled) states. Indeed, there are important differences between these two sets. For instance, the set of separable states is convex, while the set of classical states is not~\cite{LL08}. Also, measures of non-classicality such as discord do not satisfy monogamy relations~\cite{SAPB12}, which constitutes a basic property of quantum entanglement. It is usually assumed that classical states do not provide resources for information processing or information transmission, based on quantum correlations. Consider two parties $A$ and $B$, that share a quantum state of a bipartite system consisting of two subsystems $a$ and $b$ ($A$ is in possession of subsystem $a$ while $B$ is in possession of subsystem $b$). Now, assume that one or both subsystems ($a$ and $b$) are themselves composite systems. Then, as it was shown by Li and Luo, tracing over part of one (or both) subsystems $a$ and $b$, it is possible to obtain a state with quantum correlations, even if the original joint state of the composite $ab$ was classical. This is an interesting effect, because it indicates that the aforementioned classical state of the composite system $AB$ may have some ``hidden" quantum correlations. The aim of the present contribution is to study in detail this effect for some families of states of systems of three qubits.

The approach to quantum correlations proposed by Luo~\cite{Luo08} on the basis of measurement induced disturbances has two desirable features. First, it has a direct and intuitive interpretation in terms of the basic notion that in classical settings one can do a measurement on a system without disturbing it. In quantum scenarios, on the contrary, measurements usually lead to disturbances on the systems being measured. Luo applies these ideas to the study of correlations in bipartite systems. According to this approach, a bipartite system has only classical correlations if it is possible to conduct local measurements on both subsystems that do not disturb the global state of the composite system. If this can not be done, the (minimum) size of the disturbance due to local measurements constitutes a quantitative measure of the quantumness of the correlations exhibited by the system under consideration. Another advantage of Luo's proposal is that the concomitant measure of the quantum character of correlations is computationally more tractable than other measures, such as quantum discord. It is important to emphasize that both quantum discord and the notion of quantum correlations based upon measurement induced disturbances determine the same family of classical states of a quantum bipartite system. As already mentioned, these states are those described by density matrices that are diagonal in a product basis. Indeed, it is shown in~\cite{Luo08} that a quantum state $\rho$ of a bipartite system is undisturbed by appropriate (un-read) local measurements if and only if $\rho$ is diagonal in a product basis. This suggests a natural way of assessing the ``amount of quantumness'' exhibited by the correlations present in a quantum state $\rho$, by recourse to the minimum possible ``distance'' between $\rho$ and the disturbed state ${\Pi(\rho)}$ resulting from a local measurement~\cite{Luo08}.

\section{Non-classicality indicators based on measurement induced disturbances}
Given a bipartite system's density matrix ${\rho^{ab}}$, with ${\rho^a:=\tr_b{\rho^{ab}}}$ and ${\rho^b:=\tr_a{\rho^{ab}}}$ the pertinent reduced densities, one defines the measure
    \begin{equation} \label{eq:mid}
        \midi(a,b) := \mathcal{I}(a,b) - I_C(a,b) \,,
    \end{equation}
where ${\mathcal{I}(a,b):=S(a)+S(b)-S(a,b)}$ is the mutual quantum information between the two parties  $a$-$b$ of $\rho^{ab}$ and ${I_C(a,b)}$ is the classical mutual information ascribed to the post-measurement state ${\Pi[\rho^{ab}]:=\sum_{m,n}{\Pi_{mn}\rho^{ab} \Pi_{mn}}}$, such that ${\{\Pi_{mn}\}=\{\Pi_m^a\otimes\Pi_n^b\}}$ is a projective measurement (complete and bi-local) on ${\rho^{ab}}$. $S(\rho)=-\tr(\rho \log \rho)$ is the von Neumann entropy of the state $\rho$.

We are particularly interested in Measurement Induced Disturbances (MIDs). Now the set ${\{\Pi_m^a\}}$ (${\{\Pi_n^b\}}$) corresponds to  the eigen-projectors of the spectral decomposition of the state ${\rho^a}$ (${\rho^b}$). Since MIDs do not involve any kind of optimization, they may overestimate sometimes quantum correlations. This problem has been dealt with in several ways. One of them is the symmetric discord
    \begin{equation} \label{eq:sym_discord}
        \mathcal{M_S}(a,b) := \inf_{\{E_a\otimes E_b\}} \{ \mathcal{I}(a,b) - I'(a,b) \} \,,
    \end{equation}
where ${I'(a,b)}$ corresponds to the post-measurement state resulting from the general local measure ${\{E_a\otimes~E_b\}}$~\cite{WPM09,MV12,GPA11}.

Our main goal here is to detect non-classicality. As a consequence, overestimation does not constitute an important problem for us. Thus, we focus attention on MIDs given the tractability of the associated computational problem, both from the analytical and the numerical viewpoints.

As mentioned before, a bipartite state is \textit{classical} if and only if it is  diagonal in one special basis, that of the local eigen-projectors, and can thus be expressed as
    \begin{equation} \label{eq:classicalstate}
        \rho^{ab} = \sum_{m,n}{ p_{mn} \Pi_{m}^a \otimes \Pi_n^b } \,,
    \end{equation}
with ${\{p_{mn}\}}$ a bivariate probability distribution ${p_{mn}\geq0}$ and ${\sum_{m,n}p_{mn}=1}$.

\section{Quantum Correlations from Classical States}
Li and Luo~\cite{LL08} demonstrated that any separable state (classical or not) can be regarded as embedded in a classical state of a system of larger dimensionality. Reciprocally, given a classical state ${\rho^{ab}}$, any reduction would lead to a  separable state in a space of lesser dimension. Thus, a classical state's reduction might generally be a non-classical one. This is the fact on which we will concentrate our efforts, i.e., \emph{the possibility of finding reduced states correlated (in quantum fashion), starting from a classical state}.

We consider a classical state ${\rho^{ab}}$ and analyze the possibility of encountering non-classical reductions. We begin by enumerating some MID-properties that apply for an arbitrary reduced state of ${\rho^{ab}}$ whenever this classical state is defined via Eq.~\eqref{eq:classicalstate}. Assume that both parties are amenable to further decomposition  such that their associated Hilbert spaces can be cast as tensor products ${\hilb^a=\bigotimes_m\hilb^{a_m}}$ and
${\hilb^b=\bigotimes_n\hilb^{b_n}}$. The joint state of the parties $a_i$ and $b_j$ is
    \begin{equation} \label{eq:reduction}
        \rho^{a_ib_j} = \sum_{m,n}{  p_{mn} \rho_{m}^{a_i} \otimes \rho_n^{b_j} } \,,
    \end{equation}
where ${\rho_{m}^{a_i}:=\tr_{\{a_k,k\neq i\}}{\Pi_m^a}}$ and ${\rho_{n}^{b_j}:=\tr_{\{b_k,k\neq j\}}{\Pi_n^b}}$. Thus, we can apply our quantum correlations' measure between these two components:
    \begin{equation} \label{eq:midcomp}
        \midi(a_i,b_j) :=  \mathcal{I}(a_i,b_j) - I_C(a_i,b_j) \,,
    \end{equation}
so as to compute quantum correlations using Eq.~\eqref{eq:reduction}. Firstly, we realize that for any state ${\rho^{ab}}$ (classical or not), given the positivity of ${I_C(a,b)}$, one has
    \begin{equation} \label{eq:prop1}
        \midi(a,b) \leq \mathcal{I}(a,b) \,,
    \end{equation}
with strict equality ${\midi(a,b)=\mathcal{I}(a,b)}$ iff the post-measurement classical state is a product state, so that ${\Pi[\rho^{ab}]\equiv\rho^a\otimes\rho^b}$, with ${\rho^a}$ and ${\rho^b}$ coinciding with the states of the parties before measurement, that does not modifies the reduced states of $a$ and $b$.

\subsection{Some interesting bounds}
We introduce now, with regards to the reduction of ${\rho^{a_i,b_j}}$, a series of bounds that improve on Eq.~\eqref{eq:prop1}.
    \begin{itemize}
    \item Given the strong subadditivity of  von Neumann's entropy, one can verify that
        \begin{align} \label{eq:prop2}
            \midi(a_i,b_j) & \leq \mathcal{I}(a_i,b_j) \notag \\
                & \leq \text{min}\{\mathcal{I}(a_i,b),\mathcal{I}(a,b_j)\} \notag \\
                & \leq \mathcal{I}(a,b).
        \end{align}
In particular, if ${\rho^{ab}}$ is classical, then ${\mathcal{I}(a,b)}$ measures its classical correlations. Accordingly, Eq.~\eqref{eq:prop2} implies that \emph{quantum correlations between $a_i$ and $b_j$ have as an upper bound the classical correlations between $a$ and $b$ for the composite system}. Equality ${\midi(a_i,b_j)=\mathcal{I}(a_i,b_j)}$ holds iff ${\Pi[\rho^{a_ib_j}]\equiv\rho^{a_i}\otimes\rho^{b_j}}$. Moreover, if $\rho^{ab}$ is classical, then
${\mathcal{I}(a,b)\leq\min\{H(a),H(b)\}}$, with $H(a)$ ($H(b)$) the Shannon entropies of the marginal distributions $p_i^a:=\sum_j{p_{ij}}$ and $p_j^b:=\sum_i{p_{ij}}$, associated respectively to $\rho^a$ and $\rho^b$. Thus, given $\rho^{ab}$ classical, one has
        \begin{equation} \label{eq:prop2b}
            \midi(a_i,b_j) \leq \mathcal{I}(a,b) \leq \min\{H(a),H(b)\} \,.
        \end{equation}
    \item Also, ${\midi(a_i,b_j)}$ have as an upper bound the entropies of the pertinent parties, i.e.
        \begin{align} \label{eq:prop3}
            \midi(a_i,b_j) & \leq \min\{S(a_i),S(b_j)\} ,
        \end{align}
where $S(a_i)$ ($S(b_j)$) stands for von Neumann's entropy for $\rho^{a_i}$ ($\rho^{b_j}$). To prove \eqref{eq:prop3} it is enough to point out that ${S(a_i,b_j)-S(a_i)}$ and ${S(a_i,b_j)-S(b_j)}$ are concave functions because ${\rho^{a_i,b_j}}$ is separable~\cite{NK01}. Accordingly, ${S(a_i,b_j)}$ is greater
than either $S(a_i)$ or $S(b_j)$, so that ${\mathcal{I}(a_i,b_j)\leq\min\{S(a_i),S(b_j)\}}$. In view of \eqref{eq:prop2}, the bound \eqref{eq:prop3} follows immediately. This inequality reveals just how the quantum  correlations between two system's components are  conditioned by the dimensionality of the parties, with ${S(u)\leq\log[\dim(u)]}$.
    \item Lastly, note that a sufficient condition for the reduction ${\rho^{a_ib_j}}$ (of the classical state $\rho^{ab}$) to be classical as well, i.e., that
${\midi(a_i,b_j)=0}$, is that ${\{\rho_m^{a_i}\}}$ and ${\{\rho_n^{b_j}\}}$ be sets of  mutually commuting operators. In such a case, there exist common  basis of eigen-projectors for each party, ${\{\Pi_u^{a_i}\}}$ and ${\{\Pi_v^{b_j}\}}$, so that it is possible to express the composite state in the fashion
        \begin{equation} \label{eq:conmut}
            \rho^{a_ib_j} = \sum_{u,v}{ p_{uv} \Pi_u^{a_i} \otimes \Pi_v^{b_j} } \,.
        \end{equation}
It is worth mentioning, however, that the commutativity condition of the sets ${\{\rho_m^{a_i}\}}$ and ${\{\rho_n^{b_j}\}}$ is not necessary for the classicality of ${\rho^{a_ib_j}}$. It is still possible to encounter classical states even if this commutativity is not verified~\cite{LL08}.
    \end{itemize}

We pass now to a single example in which to observe the phenomenon we are interested in: a three-qubits bipartite system.

\section{Nonzero MID in classical bipartite states of  three qubits}
Consider the special bipartite state $\rho^{ab}$ exhibiting the following features. Party $a$ is comprised of two qubits while party $b$ consists of just one qubit. The state is classical for (Cf. Eq.~\eqref{eq:classicalstate})
    \begin{equation} \label{eq:3qclassical}
        \rho^{ab} = \sum_{m=1}^{4}\sum_{n=1}^{2}{  p_{mn} \Pi_{m}^{a} \otimes \Pi_n^{b} } \,,
    \end{equation}
with ${\{\Pi_{m}^{a}\equiv\Pi_{m}^{12}\}}$ the set of eigen-projectors of ${\rho^a\equiv\rho^{12}=\tr_b\rho^{ab}}$ and ${\{\Pi_{n}^{b}\equiv\Pi_{n}^{3}\}}$ that of $\rho^3$.

\begin{figure}
        \centering
        \includegraphics[width=.8\columnwidth]{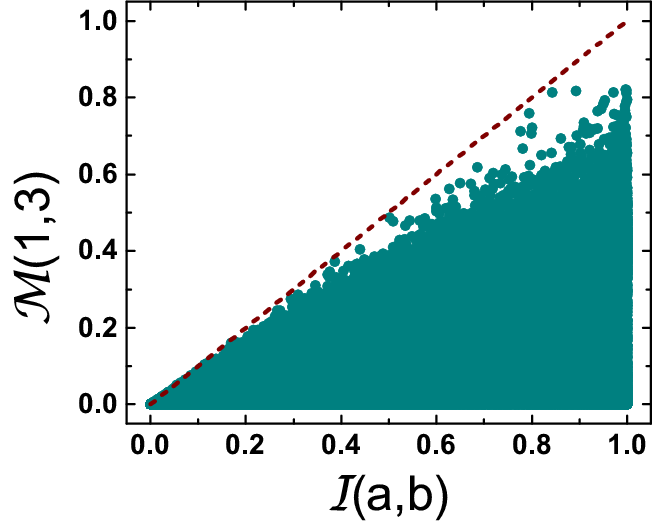}
        \caption{MID corresponding to subsystem $(1,3)$ vs. mutual information for the classical state  $ab$, evaluated for ${\sim10^5}$ randomly selected states. The dotted line (red) depicts the bound ${\midi(1,3)\leq\mathcal{I}(a,b)}$.}
        \label{fig:MvsI_rand}
    \end{figure}

So as to encounter quantum correlations in the reduced state ${\rho^{13}=\tr_2{\rho^{ab}}}$, we need that some of the members of the set ${\{\rho_m^1=\tr_2\Pi_m^a\}}$ do not commute amongst themselves. For instance, define the operators ${\Pi_m^a:=\ket{a_m}\bra{a_m}}$ with
    \begin{equation} \label{eq:a_base}
        \begin{cases}
        \ket{a_1} = \ket{00}\,, &  \ket{a_2} = \ket{10} \,, \\
        \ket{a_3} = \ket{+1}\,, & \ket{a_4} = \ket{-1} \,,
        \end{cases}
    \end{equation}
with the states given by ${\ket{+}=(1/\sqrt{2})(\ket{0}+\ket{1})}$ and ${\ket{-}=(1/\sqrt{2})(\ket{0}-\ket{1})}$, for the basis of $a$, and the operators ${\Pi_n^b:=\ket{b_n}\bra{b_n}}$ with ${\ket{b_1}=\ket{0}}$ and ${\ket{b_2}=\ket{1}}$ for subsystem $b$. Using these basis, we numerically compute the measures $\midi(1,3)$ for a sample of $10^4$ states with randomly generated  distributions ${\{p_{mn}\}}$. In the graph $\midi(1,3)$ vs. ${\mathcal{I}(a,b)}$ (Fig.~\ref{fig:MvsI_rand}) we see that the ensuing states almost completely fill up the region  lying under the straight line of unit slope ${\midi(1,3)=\mathcal{I}(a,b)}$. Such result agrees with the upper bound  anticipated by  Eqs.~\eqref{eq:prop2} and \eqref{eq:prop3}. Note that all classical states of the composite system ${a-b}$ (not only those belonging to the family \eqref{eq:3qclassical}-\eqref{eq:a_base}) correspond to points that must lie within the above triangular region of the ${\mathcal{I}(a,b)-\midi(1,3)}$ plane.

We consider now states lying on the border of the region depicted in Fig.~\ref{fig:MvsI_rand}. For the lower border we have ${\{\midi(1,3)=0;\,0\leq\mathcal{I}(a,b)\leq1\}}$, for the right-side one ${\{\mathcal{I}(a,b)=1;\,0\leq\midi(1,3)\leq1\}}$, and for the upper border ${\{\midi(1,3)=\max\midi(1,3)|_{\mathcal{I}(a,b)}\}}$. We shall provide parameterized families of states that correspond to the above borders, in order to illustrate the fact that these frontiers can be actually reached by classical states of the ${a-b}$ composite system. If the contrary is not explicitly stated, we are going to consider states belonging to the family \eqref{eq:3qclassical}-\eqref{eq:a_base}.

\begin{enumerate}
    \item \emph{${\midi(1,3)=0}$ Border}. It is easy to construct classical states $\rho^{ab}$ such that  $\rho^{13}$ is classical as well. Any state defined using the basis ${\{\Pi_m^a\}}$ with commuting operators from the set ${\{\rho_m^1=\tr_2\Pi_m^a\}}$ will do. It is convenient to have a free parameter at our disposal so as to have the classical mutual information to traverse the interval ${0\leq\mathcal{I}(a,b)\leq1}$. For example, we have the one-parameter family ${\rho^{ab}_\alpha=\alpha\ket{000}\bra{000}+(1-\alpha)\ket{101}\bra{101}}$,
with ${0\leq\alpha\leq1}$. Then, for this family, we have ${\mathcal{I}^\alpha(a,b)=-\alpha\log\alpha-(1-\alpha)\log(1-\alpha)}$, ${\midi^\alpha(1,3)=0}$ that yields the whole lower border of the region we are interested in (see Fig.~\ref{fig:M0border}).
    \begin{figure}
        \centering
        \includegraphics[width=.8\columnwidth]{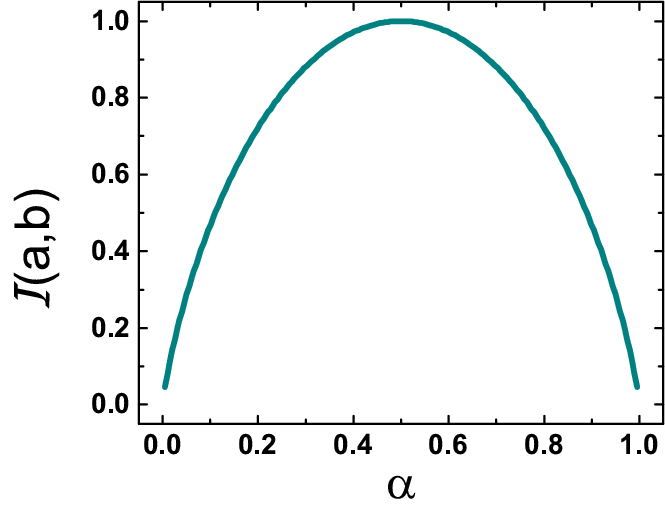}
        \caption{Mutual information as a function of the parameter $\alpha$ for the family ${\rho^{ab}_\alpha}$ that reproduces the lower border of the regi\'on for which ${\midi(1,3)=0}$.}
        \label{fig:M0border}
    \end{figure}
    \item \emph{${\mathcal{I}(a,b)=1}$ Border}. Here we need a family of states such that i) the reduction $\rho^{13}$ exhibit quantum correlations and also ii) maximizes the mutual information. For this second condition to hold we need a strong correlation, like that given by ${p_{ij}=p_i^a\delta_{ij}}$. Accordingly,
${\mathcal{I}(a,b)=S(a)=S(b)=S(a,b)}$, that is maximized by uniform marginal distributions. The states-family
    \begin{equation} \label{landafam}
        \rho^{ab}_\gamma=\frac{1}{2}\ket{000}\bra{000}+\frac{1}{2}\ket{\psi_\gamma 1 1}\bra{\psi_\gamma 1 1} \,,
    \end{equation}
with ${\ket{\psi_\gamma}=\cos\gamma\ket{0}+\sin\gamma\ket{1}}$ and ${0\leq\gamma\leq\pi}$, foots the bill and also satisfies the condition that ${\rho_1^1:=\tr_2\Pi_1^a=\ket{0}\bra{0}}$ and ${\rho_2^1:=\tr_2\Pi_2^a=\ket{\psi_\gamma}\bra{\psi_\gamma}}$ do not generally commute. This raises hopes of ending up with ${\midi(1,3)\neq0}$. Indeed, one can analytically find the MID for this family. The eigenvalues of ${\Pi[\rho^{13}_\gamma]}$ are ${(1/4)(1\pm\cos\gamma)}$, both exhibiting a twofold degeneracy. We see in Fig.~\ref{fig:rightborder_MvsAng} that this family yields the right-side border that concerns us here, with
${0\leq\midi(1,3)\leq1}$. The states ${\rho^{ab}_\gamma}$ do not belong to the family \eqref{eq:3qclassical}-\eqref{eq:a_base}, but they are nevertheless classical states of the composite system ${a-b}$ illustrating that the frontier ${\mathcal{I}(a,b)=1}$ (with values of $\midi(1,3)$ covering the full range ${[0,1]}$) can be reached by these kind of states.
    \begin{figure}
    \centering
    \includegraphics[width=.8\columnwidth]{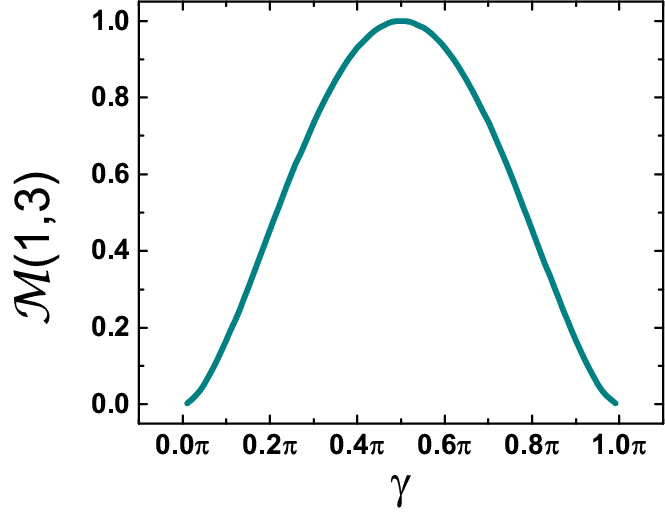}
    \caption{MID for the subsystem ${(1,3)}$, as a function of the angle $\gamma$, for the family  ${\rho^{ab}_\gamma}$, that reproduces the right-side border of the triangular region of Fig.~\ref{fig:MvsI_rand}.}
    \label{fig:rightborder_MvsAng}
    \end{figure}
    \item \emph{Upper border.} We were unable to find a family of states maximizing MID for all values of ${\mathcal{I}(a,b)}$ within the interval that interest us. We did encounter a class of one-parameter states that reach the maximum possible value ${\midi^{max}_{(1,3)}=1}$. Fixing ${p_{11}=1-2\lambda}$ and ${p_{31}=\lambda=p_{42}}$ in the above mentioned basis (Cf. Eq.~\ref{eq:a_base}) we obtain the family
    \begin{align} \label{eq:M13max}
    \rho^{ab}_\lambda & = (1-2\lambda)\ket{000}\bra{000} \notag\\
            & + \lambda(\ket{+10}\bra{+10}+\ket{-11}\bra{-11}).
    \end{align}
We shall consider the range of $\lambda$-values ${[0,1/2)}$. Within this range we have that the mutual information is ${\mathcal{I}_\lambda(a,b)=-\lambda\log\lambda-(1-\lambda)\log(1-\lambda)}$. The eigenvalues of ${\rho^{13}_\lambda}$ are ${\{0,\lambda,\frac{1}{2}(1-\lambda+c_\lambda),\frac{1}{2}(1-\lambda-c_\lambda)\}}$, with ${c_\lambda:=\sqrt{1-4\lambda+5\lambda^2}}$. Those for the post-measurement ${\Pi[\rho^{13}_\lambda]}$ state are ${\{1-\frac{3}{2}\lambda,\frac{\lambda}{2},\frac{\lambda}{2},\frac{\lambda}{2}\}}$. Accordingly, the associated entropies become
    \begin{align} \label{eq:S13}
    S(1&,3) = -\lambda\log\lambda \notag\\
    & - \frac{1}{2}(1-\lambda+c_\lambda)\}\log(\frac{1}{2}(1-\lambda+c_\lambda)\}) \notag\\
    & - \frac{1}{2}(1-\lambda-c_\lambda)\}\log(\frac{1}{2}(1-\lambda-c_\lambda)\}) ,
    \end{align}
and
    \begin{equation} \label{eq:S13m}
    S'(1,3) = 1 -\frac{2-3\lambda}{2}\log(2-3\lambda) -\frac{3}{2}\lambda\log\lambda \,.
    \end{equation}
Finally, from \eqref{eq:S13}-\eqref{eq:S13m} and making ${\midi^\lambda(1,3)=S'(1,3)-S(1,3)}$ one obtains the MID for this class of one-parameter states. Fig.~\ref{fig:MyAvsI_f4} depicts the pertinent results for the family ${\rho^{ab}_\lambda}$. For verification purposes, we also evaluated in numerical fashion the optimized measure ${\mathcal{M_S}(1,3)}$.
    \begin{figure}
        \centering
        \includegraphics[width=.8\columnwidth]{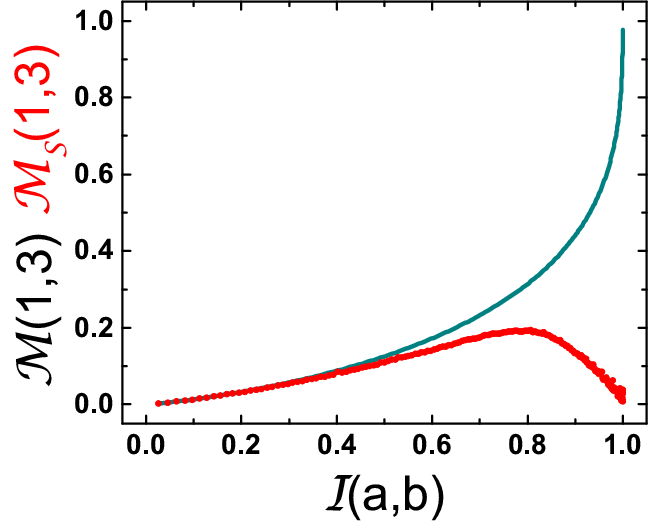}
        \caption{$\midi(1,3)$ (cyan)  and  ${\mathcal{M_S}(1,3)}$ (red) for the subsystem ${(1,3)}$ vs. the mutual information of the classical state $ab$ corresponding to the family ${\rho^{ab}_\lambda}$.}
        \label{fig:MyAvsI_f4}
    \end{figure}

A measure optimized over all local projective measurements significantly differs from the MID evaluated from ${\mathcal{I}(a,b)\approx0.5}$ (Fig.~\ref{fig:MyAvsI_f4}). This evidences the MID-overestimation of quantum correlations. However, it is clear that the two measures agree on {\it which} are the states that exhibit quantum correlations.

\vfill

\end{enumerate}

\section{Conclusions}
We have  investigated in some detail the precise way in which, starting from a classically correlated state, one finds by reduction a state  correlated in quantum fashion.

Things were illustrated with reference to the system of lowest dimensionality where this effect can take place, that is, a three-qubit system. In this case we encountered a relationship between  the classical mutual information of the composite state and the maximum quantum correlation for the reduced state. We also found  families of classical states that actually reach the bounds foreseen by such relationship.

Our results may be of interest because they provide  a better insight on the relation between classical and quantum correlations. In particular, the fact that (separable) quantum-correlated states can be obtained as reductions of classical states raises some interesting questions concerning the status of non-entangled
quantum correlated states as resources for information-related tasks. In this regards, it is instructive to consider the following scenario. Consider a bipartite quantum system consisting of subsystems $a$ and $b$. Assume that the Hilbert spaces of $a$ and $b$ have dimensions that are composite numbers (that is, they are not prime numbers). In that case the subsystems $a$ and $b$ can always be regarded formally as composite systems (this possibility is related to the fact that the tensor-product structure of the Hilbert space of composite systems is indeed observe-dependent. See for instance~\cite{ZLL04} and references therein). Consequently, a classically correlated state of the composite $ab$ can have ``hidden" quantum correlations that correspond to a reduced state obtained by tracing over ``part" of subsystems $a$ and $b$, if these subsystems are appropriately regarded as factorized into subsystems. This, in turn, suggests the intriguing possibility that, besides the quantum correlations themselves, the dimensionality of the subsystems $a$ and $b$ should be regarded as ``resources" in the sense that, the larger are these dimensions (provided that they are composite numbers) the larger the amount of ``hidden" quantum correlations that can be extracted from the original state by recourse to appropriate reductions. We plan to address this point in a forthcoming contribution.



\begin{thebibliography}{}

\bibitem{OZ01} H. Ollivier, W.H. Zurek, Phys. Rev. Lett. {\bf 88}, 017901 (2001).
\bibitem{HV01} L. Henderson, V. Vedral, J. Phys. A {\bf 34}, 6899 (2001).
\bibitem{Vedral03} V. Vedral, Phys. Rev. Lett. {\bf 90}, 050401 (2003).
\bibitem{LC10} M.D. Lang, C.M. Caves, Phys. Rev. Lett. {\bf 105}, 150501 (2010).
\bibitem{Luo08} S. Luo, Phys. Rev. A {\bf 77}, 022301 (2008).
\bibitem{Luo08b} S. Luo, Phys. Rev. A {\bf 77}, 042303 (2008).
\bibitem{LL08} N. Li, S. Luo, Phys. Rev. A {\bf 78}, 024303 (2008).

\bibitem{MV12} K. Modi, A. Brodutch, H. Cable, T. Paterek, V. Vedral, Rev. Mod. Phys. {\bf 84}, 1655 (2012).
\bibitem{GPA11} D. Girolami, M. Paternostro, G. Adesso, J. Phys. A {\bf 44}, 352002 (2011).

\bibitem{LLZ07} N. Li, S. Luo, Z. Zhang, J. Phys. A {\bf 40}, 11361 (2007).
\bibitem{DVB10} B. Dakic, V. Vedral, C. Brukner, Phys. Rev. Lett. {\bf 105}, 190502 (2010).
\bibitem{RRA2011} L. Roa, J.C. Retamal, M. Alid-Vaccarezza1,
Phys. Rev. Lett. {\bf 107}, 080401 (2011).
\bibitem{MPP12} A.P. Majtey, A.R. Plastino, A. Plastino, Physica A {\bf 391}, 2491 (2012).





\bibitem{SRN08} A. SaiToh, R. Rahimi, M. Nakahara, Phys. Rev. A {\bf 77}, 052101 (2008).
\bibitem{RS10} R. Rahimi, A. SaiToh, Phys. Rev. A {\bf 82}, 022314 (2010).

\bibitem{WPM09} S. Wu, U.V. Poulsen, K. M\o lmer, Phys. Rev. A {\bf 80}, 032319 (2009).
\bibitem{AD10} A. Auyuanet, L. Davidovich, Phys. Rev. A {\bf 82}, 032112 (2010).



\bibitem{OHHH02} J. Oppenheim, M. Horodecki, P. Horodecki, R. Horodecki, Phys. Rev. Lett. {\bf 89}, 180402 (2002).
\bibitem{LF11} S. Luo, S. Fu, Phys. Rev. Lett. {\bf 106}, 120401 (2011).

\bibitem{RCC10} R. Rossignoli, N. Canosa, L. Ciliberti,
Phys. Rev. A {\bf 82}, 052342 (2010).

\bibitem{SAPB12}
A. Streltsov, G. Adesso, M. Piani, and D. Bruss, Phys. Rev. Lett.
{\bf 109}, 050503 (2012).

\bibitem{NK01} M. A. Nielsen, J. Kempe, Phys. Rev. Lett. {\bf 86}, 5184 (2001).

\bibitem{ZLL04} P. Zanardi, D.A. Lidar, and S. Lloyd, Phys. Rev. Lett. {\bf 92}, 060402 (2004).


\end{thebibliography}
\end{document}